\documentclass[preprint,superscriptaddress,amsmath,amssymb,12pt]{article}

\usepackage{graphicx}
\usepackage{dcolumn}
\usepackage{bm}
\usepackage{setspace}
\begin{document}

\title{ Sublattice asymmetry of impurity doping in graphene: A review}

\author{James A. Lawlor \\ School of Physics, Trinity College Dublin, Dublin 2, Ireland
\and Mauro S. Ferreira \\ School of Physics, Trinity College Dublin, Dublin 2, Ireland \\ CRANN, Trinity College Dublin, Dublin 2, Ireland}

\date{\today}

\maketitle

\begin{abstract}

In this review we highlight recent theoretical and experimental work on sublattice asymmetric doping of impurities in graphene, with a focus on substitutional Nitrogen dopants.
It is well known that one current limitation of graphene in regards to its use in electronics is that in its ordinary state it exhibits no band gap.
By doping one of its two sublattices preferentially it is possible to not only open such a gap, which can furthermore be tuned through control of the dopant concentration, 
but in theory produce quasi-ballistic transport of electrons in the undoped sublattice,
 both important qualities for any graphene device to be used competetively in future technology. 
We outline current experimental techniques for synthesis of such graphene monolayers and detail theoretical efforts to explain the mechanisms responsible for the effect,
before suggesting future research directions in this nascent field.
\end{abstract}
 
\section{Review}

\subsection{Introduction}

With its excellent transport properties and low dimensionality, graphene, an atomically thin layer of Carbon atoms bonded together in a hexagonal lattice,
initially seems a strong candidate for use in many future commercial applications such as ultra high-speed transistors, integrated
circuits and other novel devices \cite{geim_rise_2007, novoselov_roadmap_2012, chen_intrinsic_2008}.
One of the main problems with using regular graphene for such applications is the absence of a band gap in the electronic band structure \cite{allen_honeycomb_2010}, and as a result 
any Field Effect Transistors (FETs) made using the material (so-called GFETs) would be unable to be switched off, rendering it useless as a logic device \cite{schwierz_graphene_2010, lin_operation_2009,zhang_direct_2009}.
The most natural way to approach this issue is therefore to introduce a sizeable band gap and hence allowing more control over the current flow.

Although several methods exist to induce a band gap, for example 1D quantum confinement by construction of graphene nanoribbons (GNRs) \cite{li_chemically_2008},
stacking of monolayers with perpendicular electric fields \cite{lin_operation_2009,zhang_direct_2009},
strain \cite{naumov_gap_2011} and mounting on a substrate
\cite{novoselov_graphene:_2007, zhou_substrate-induced_2007, chang_band_2013},
these methods are not without problems.
Alternatives are therefore sought after with the minimum standard to
meet or exceed the limits of silicon semiconductor technology, which is characterised by a current on/off ratio of roughly $10^4$-$10^7$ and a band
gap of at least $340 meV$ whilst maintaining high carrier mobility \cite{avouris_graphene:_2010,schwierz_graphene_2010, kim_role_2011}.

Alteration of the crystal structure through the introduction of foreign dopants is one of the more realistic avenues of approach in realising this goal.
Atomic dopants like Boron (B) or Nitrogen (N) are a similar size to Carbon and can be introduced easily in a variety of graphene growth processes,
typically replacing Carbon sites and forming substitutional impurities in the lattice, positively (p-) and negatively (n-) doping the system for B and N dopants respectively \cite{islam_effect_2014}.
 Early theoretical attempts at investigating the electronic properties of such a material found that a periodic arrangement of B or N
dopants, forming a dopant superlattice, would open a band gap \cite{casolo_band_2011},
but that a random distribution of dopants among lattice sites yields no band gap \cite{lherbier_charge_2008}.
 Further investigations using vacancies, where carbon atoms are removed from the lattice, found that both superlattices \cite{martinazzo_symmetry-induced_2010}
 and random distributions restricted to one of the two graphene sublattices \cite{palacios_vacancy-induced_2008} both lead to a tunable band gap,
 and in the latter case an emergence of magnetic properties in the system \cite{cresti_broken_2013,pereira_modeling_2008}.
 Whilst these findings are certainly interesting, their scalability, and hence commercial application, is prohibited by the standard of precision that must be met in order to produce such materials.
  Circumventing these problems using Nitrogen doped graphene has only come about relatively recently which we detail in the following section.

\subsection{A Brief History of Nitrogen Doped Graphene}
 
Exploiting the effects of Nitrogen dopants on the transport properties of graphene has been an interesting experimental research topic for the last 5 years \cite{wei_synthesis_2009},
whereby the methods and techniques to achieve this have been developed to include Chemical Vapour Deposition (CVD), using$NH_3$as a precursor,
arc discharge \cite{guan_preparation_2011}, embedded nitrogen and carbon sources within a metal substrate \cite{zhang2011synthesis}, ion implantation \cite{doi:10.1021/nn502438k, doi:10.1021/nl402812y}, Ammonia \cite{lin_ammonia_2010} or Nitrogen plasma \cite{wang_nitrogen-doped_2010,joucken_localized_2012} treatments, and capable of achieving Nitrogen dopant
concentrations of up to around $10$\% \cite{wang_review_2012} and with direct applicablity to GFET technology and bio-sensing \cite{wang2010nitrogen}.
Although CVD is one of the more challenging methods, it seems the most reliable option and yields the best quality Nitrogen doped graphene sheets \cite{wang_strategies_2014} and
single continuous sheets can be synthesised on the centimeter scale  \cite{gao_simple_2012}.
Using Nitrogen dopants alone through CVD can yield bandgaps up to 200meV \cite{usachov_nitrogen-doped_2011} and by inclusion of Boron co-dopants, through a tailored growth process, this can be expanded to
around 600meV with 6\% total dopant concentration \cite{chang_band_2013}. 
Boron doping alone has shown to also open a band gap, tunable with dopant concentration \cite{doi:10.1021/nn3005262, fan2012band}.
This method results in small B-N domains embedded in the graphene sheet, so the precise mechanism behind the band gap opening may not be the same as that seen in a random B-N ensemble. 
Although these gaps are quite sizeable, the effect of scattering would lead to a detrimental impact on the transport characteristics.
However, the recent theoretical prediction of sublattice asymmetry, a situation where dopants are distributed in one sublattice only, is expected to yield very low scattering transport properties.
 The electron wavefunctions in such a system are shown to mainly exist on the sublattice without dopants and can travel almost unhindered \cite{lherbier_electronic_2013, botellomendez_modeling_2013}, 
 which is very promising for overcoming the scattering problem.
 
\begin{figure}
    \centering
 \includegraphics[scale=0.5]{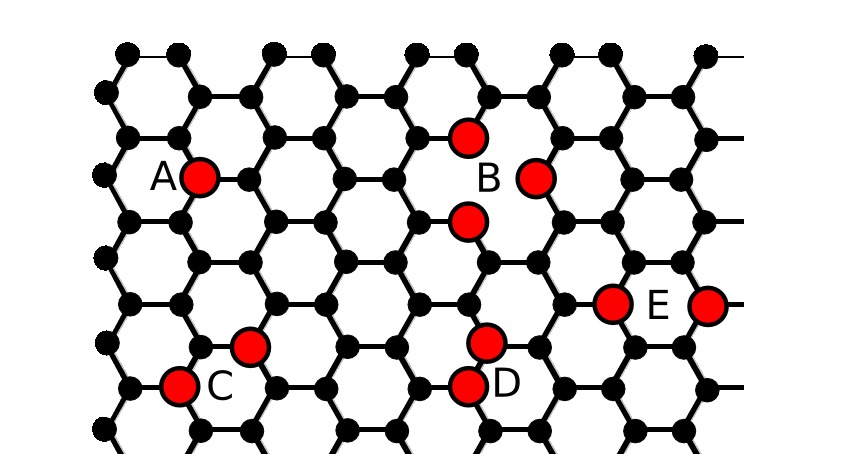}
  \caption{Schematic of a graphene lattice with the most common experimentally observed species of substitutional Nitrogen dopants:
  (A) a single Graphitic, (B) three Pyridinics, (C) one $N_2^{AA}$ pair, (D) one $N_2^{AB}$ pair and (E) one $N_2^{AB'}$ pair. }
  \label{fig:schematic}
  \end{figure}

It was not until 2011 that a viable experimental approach was found by Zhao et al. \cite{zhao_visualizing_2011}.
 They discovered that graphene grown via Chemical Vapour Deposition (CVD) in the presence of ammonia ($NH_3$)
naturally incorporates Nitrogen atoms as substitutional so-called 'graphitic' dopants (see Fig. \ref{fig:schematic} A) into the crystal, and with a distinct sublattice segregation of dopants.
Indeed further research has uncovered a less pronounced asymmetry phenomenon using graphene implanted with nitrogen impurities followed by a high temperature annealing process \cite{doi:10.1021/nn502438k}, 
 and it seems reasonable that there is a common mechanism with the CVD method.
 Due to the limited published work on this implantation and annealing method, our main focus will be on CVD and it is here that we begin our main discussions.
 We begin with a section covering in-depth the experimental work done to date, followed by a section on theoretical aspects and ending with a future outlook and conclusions.

\subsection{Experimental Observation of Sublattice Asymmetry of N-graphene}

The main focus of the work by Zhao et al. was the observation and characterization of nitrogen dopants via scanning tunnel spectroscopy (STS), followed by
a short investigation on the transport properties of the resulting graphene.
 Their experimental observation of same sublattice segregation, at least on local scales, was noted as a curiosity but was not discussed in depth.
A key finding by the researchers was that by varying the ammonia precursor concentration they could adjust the resulting embedded dopant concentration. 
 Further theoretical work confirmed that a tunable band gap would be obtainable using this method \cite{lherbier_electronic_2013}, and along with the findings of Zhao et al. this 
 sparked further interest into this phenomenon.
 It should be noted that the experimental conditions were a high vacuum CVD process on a Cu (111) substrate with a temperature of $1000\,^{\circ}\mathrm{C}$, using
$CH_4$ and $NH_3$ as graphene and Nitrogen precursors respectively for 18 minutes reaction time achieving a dopant concentration of 0.3\%.

Shortly after this work a more focussed investigation was undertaken by Lv et al. \cite{lv_nitrogen-doped_2012}.
They were able to replicate the findings of the first group and synthesised large areas of N-doped monolayer graphene with the additional finding that
by increasing the pressure in the growth phase from a high vacuum to atmospheric-pressure CVD would result in an abundance of so-called $N_2^{AA}$  pairs \cite{lv_nitrogen-doped_2012} (see Fig. \ref{fig:schematic} (C)),
 which are thought to be a result of a higher number of intermolecule collisions during formation. 
 Using the same substrate and precursors, they were also able to find that a minimum reaction time of 5 minutes and temperatures of 800C were optimum for the N-doped graphene synthesis and moreover 
 it was found that singular Nitrogen dopants are more frequently found for reaction times below 10 minutes with the ratio of $N_2^{AA}$ to single N increasing with reaction time.
 The samples had a typical concentration of around 0.25\% Nitrogen dopants.

More recently the addition of Boron dopants into the lattice and their effects have been studied \cite{zhao_local_2013} using high vacuum CVD growth and adding $B_2 H_6$ diborane instead of ammonia as the dopant precursor,
achieving concentrations on the order of 0.3\%.
The researchers compared B-doped and N-doped systems in detail and found no detectable sublattice asymmetry in the case of B-doped systems and thus the dopants were distributed evenly between sublattices.
The reason for this is thought to be due to a strong interaction between the Cu (111) substrate and the Boron impurity, in contrast to a very weak interaction with the substrate for Nitrogen impurities\cite{zhao_local_2013}.

\begin{figure}
    \centering
  \includegraphics[scale=1.0]{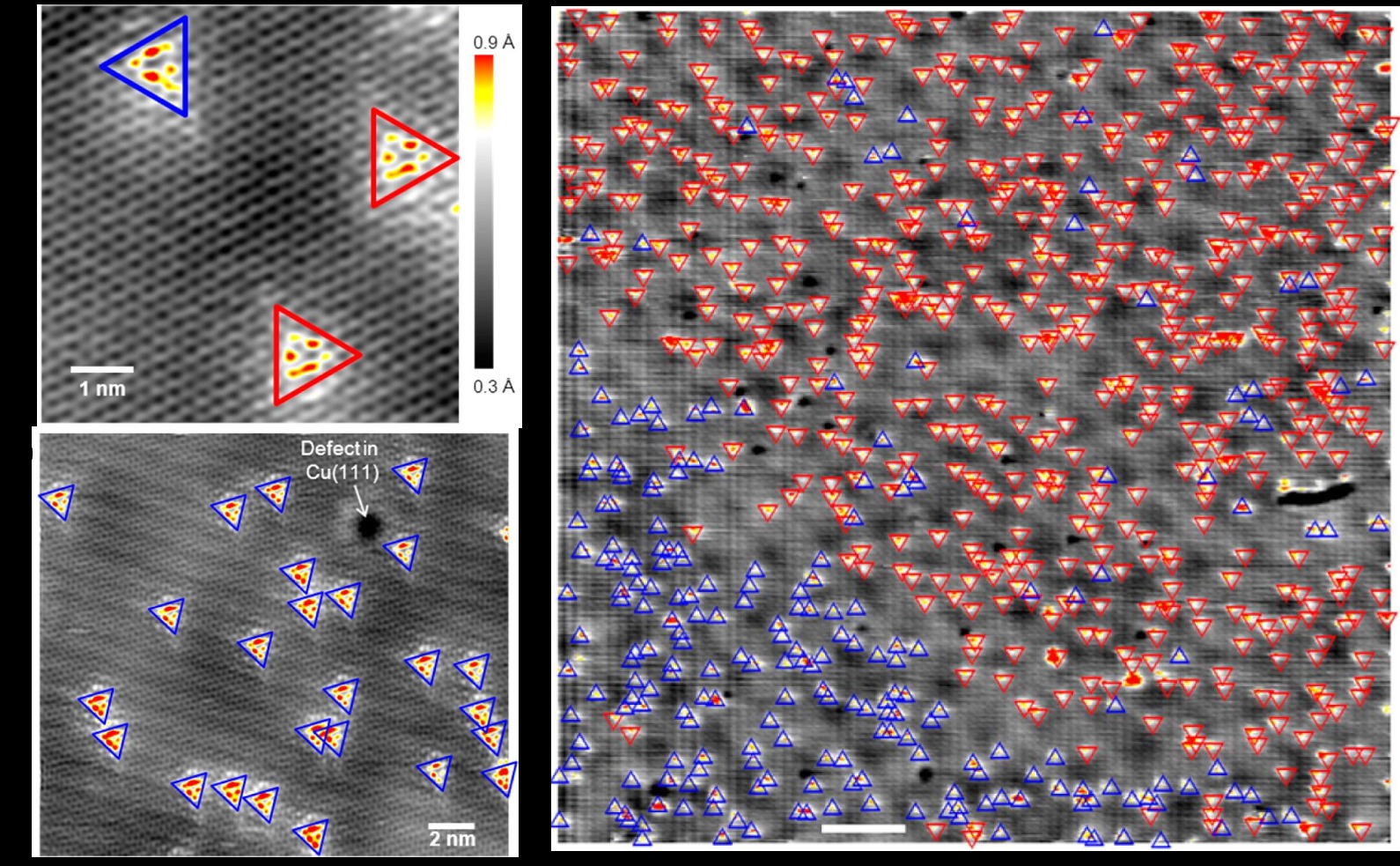}
  \caption{ STM images of nitrogen doped graphene on (a) 7nm$^2$, (b) 20nm$^2$ and (c) 100nm$^2$ scales, adapted with permission from Zabet-Khosousi et al. \cite{zabet-khosousi_segregation_2014}.
  Copyright 2014 American Chemical Society.
  The dopants locations are identified by finding bright spots on the STM image, corresponding to slight perturbations in the positions of the neighbouring carbon atoms.
  The dopant sublattice can be found through the orientation of the bright triangle features, where opposite sublattices appear as mirror images of each other as demonstrated in (a). 
  The red and blue triangles in each subfigure correspond to impurities on the different sublattices.
  Part (c) best demonstrates that although the distribution of dopants appears random, when their sublattice is identified we see two very large domains appear with opposite sublattice segregation. 
  Shown at the bottom, off-center left of (c) is a white 10nm scalebar.}
  \label{fig:zabet}
  \end{figure}

The most recent and thorough study of asymmetric Nitrogen dopants in graphene is by Zabet-Khosousi et al. \cite{zabet-khosousi_segregation_2014} who again used the CVD growth process but with a pyridine ($C_5 H_5 N$) precursor for the Nitrogen and graphene instead of the conventional ammonia/methane mix.
 Fig. \ref{fig:zabet} shows a rather striking STM image of a large area of N-doped graphene grown using this method, showing two clearly defined domains of different sublattice preference
 where the sharp borders are thought to arise from terracing of the substrate disrupting the asymmetry effect.
 The method was tailored to use a very smooth Cu (111) substrate and so they were able to achieve domain sizes far larger than previous efforts.
 Typical concentrations found with this method were around 0.2\%, very similar to the previous ammonia/methane method which points to an equivalent growth mechanism.
 By comparing N-doped graphene systems synthesised with this method to other systems made via Nitrogen ion bombardement and ammonia post-treatment of a pristine graphene sheet,
where no sublattice asymmetric configurations were found, it was suggested that the sublattice asymmetry must be occuring during formation of the graphene sheet. 
 This is supported by the recent experiments using ion bombardement followed by high temperature annealing by Telychko et al. \cite{doi:10.1021/nn502438k},
where the impurities may be reconfiguring themselves locally during cooling. This phenomenon is not seen without the annealing process \cite{doi:10.1021/nl402812y}.

 The common elements of all the aforementioned experimental reports are that they all use a Cu (111) substrate, all have reaction times exceeding 10 minutes in a temperature range of $800$-$900\,^{\circ}\mathrm{C}$,
 and all have used either ammonia or pyridine precursors to achieve concentrations of up to 0.3\%. 
A review of other methods for N-doped graphene synthesis finds no mention of sublattice asymmetry \cite{joucken_localized_2012,wang_review_2012}, 
it is known that this effect is mainly detectable through careful Scanning Tunnel Microscopy of large areas. 
 The most natural explanation for this lack of observation is then either that it is not being observed due to simply not looking,
 or that sublattice asymmetry is not a very robust effect and therefore requires a careful choice of synthesis procedure.
Research into the mechanisms behind the phenomenon is therefore paramount in answering this question and could shed light on whether other methods or even dopant species are at all possible
and this leads us naturally to a review of the current theoretical models of the effect.

\subsection{Theoretical Models of the Segregation Effect}

It is clear that any degree of asymmetry between the two equivalent sublattices is the result of a symmetry breaking operation.
Current theoretical attempts to explain the sublattice asymmetry in the Nitrogen doping seen in the experiments discussed in the previous section suggest that
the effect comes from either the energetically preferable positioning of Nitrogen on the graphene edge during the growth process \cite{zabet-khosousi_segregation_2014,deretzis_origin_2014}, where
the symmetry breaking effect is the edge structure, 
or from inter-impurity interactions in the impurity ensemble \cite{lv_nitrogen-doped_2012, lawlor_sublattice_2014} where the symmetry is broken by the impurities themselves.
These models will hereafter be referred to as the edge growth and interaction models respectively.

The edge growth model suggests that during the growth phase the most energetically favourable position for a Nitrogen dopant being incorporated into a graphene edge
would be that which leads to a same sublattice configuration for all impurities in a domain. 
Through Density Functional Theory (DFT) calculations involving a graphene nanoribbon on a Cu(111) substrate, aiming to reproduce experimental conditions, a thorough investigation into the energetic favourable
position of single graphitic Nitrogens in the GNR was undertaken and the energetic difference between placing the Nitrogen at one edge site over another on
the opposite sublattice was found to be a substantial 1.3eV.
One drawback of this method is that one would expect therefore that the clearly defined segregation domains seen experimentally (see Fig. \ref{fig:zabet} and \cite{zabet-khosousi_segregation_2014}) would 
naturally fall in with the graphene grain boundaries, but this has not been observed.
The role of inter-impurity interactions has been considered as an alternative, through both Tight Binding \cite{lawlor_sublattice_2014} and DFT \cite{lv_nitrogen-doped_2012, hou_interplay_2012} formalisms, 
although with differing conclusions.
The Tight Binding method by Lawlor et al. \cite{lawlor_sublattice_2014} was part of a more in-depth theoretical investigation, and 
suggests that a Friedel Oscillation-like perturbation in the long-range inter-impurity interactions \cite{lambin_long-range_2012} arise from adding impurities to the system,
 leading to a shift in the system's Fermi energy. 
Such a mechanism would explain both the non-commensurability between crystal grain boundaries and segregation domains, and would suggest why the alternative synthesis method of
 nitrogen implantation followed by high temperature annealing also results in impurities preferring to occupy the same sublattice \cite{doi:10.1021/nn502438k}.
Furthermore, the authors predict that as a result of this interaction and the reduction in nearest neighbour distance with increasing concentration, there exists a critical dopant concentration
beyond which no sublattice asymmetry would be observed, as the energetic minimum occurs when the impurities are distributed evenly between sublattices. 
Although the exact value of this critical concentration is dependant on how the impurity is parameterised within the tight binding regime,
 the authors predict it to lie between 0.1 \% and 0.8\%.
Comparing to the highest experimentally reported doping concentrations for samples with sublattice asymmetry (0.3\%), it is a hope that further experiments 
can clarify whether such a critical concentration exists. 
If the prediction of a critical concentration is accurate it would limit the band gap of a segregated device to around 100meV \cite{lherbier_electronic_2013}. 
This figure, however, comes from matching Tight Binding and DFT band structure results, a method which is known to systematically underestimate such band gaps. 
 Nevertheless, the band gap obtained can be expected to be much below that required for a GFET device.

 In-depth DFT calculations by Hou et al. \cite{hou_interplay_2012} found that the interactions between nitrogen impurities are generally repulsive in nature. 
More specifically, when the two impurities are placed close together the system energy is minimised for opposite sublattice configurations, with the exception of $N_2^{AB}$ .
This contradicts the experimental reports of Lv et al. \cite{lv_nitrogen-doped_2012} who found an abundance of $N_2^{AA}$ defects in their samples. 
 A simplistic Tight Binding approach predicts that two identical impurities in close proximity to one another
would indeed have an energetic minimum when both occupy the same sublattice \cite{lawlor_friedel_2013}, 
 however this method ignores coulombic interaction which will be dominant at close separations.
Interestingly, the DFT approach of Lv et al. of such Nitrogen pairs finds that the lowest energy configuration would be $N_2^{AB'}$  (see Fig \ref{fig:schematic}),
approximately 0.3eV lower than that of$N_2^{AA}$ . 
This appears to be in disagreement with experiments where$N_2^{AA}$  is more commonly found than$N_2^{AB'}$ .
As a final note their calculations have also shown that two $N_2^{AA}$  pairs have a lower energy when they share the same sublattice, however further
studies have shown that the overall energy change resulting from an $N_2^{AA}$ pair in the sheet is very high \cite{deretzis_origin_2014}.
 It is possible that these impurities also come from edge growth, but the controllable presence of $N_2^{AA}$  in nitrogen ion implanted graphene \cite{doi:10.1021/nn502438k} indicates again
 that this may not be the full picture as in this case the complete graphene sheet is already fabricated.

  \subsection{ Predicted Electronic Properties }
\begin{figure}
  \centering
  \includegraphics[scale=0.4]{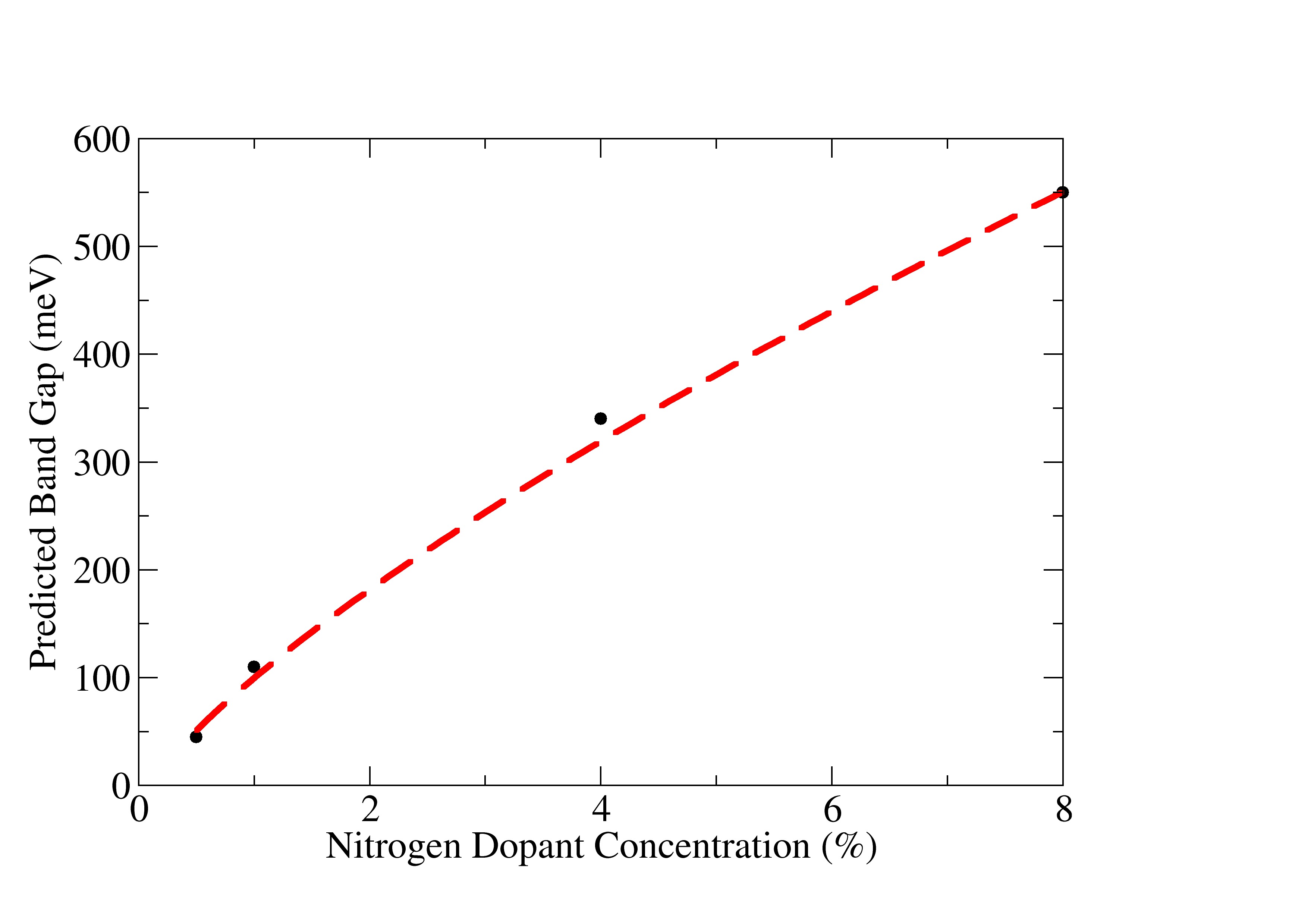}
  \caption{ Predicted band gap against Nitrogen dopant concentration.
  Black circles are calculated values from \cite{lherbier_electronic_2013} and the red dashed line shows the expected band gap scaling with concentration, according to the power $3/4$ as discussed in the text.}
  \label{fig:bandgaps}
 \end{figure}

The earliest attempts to study Nitrogen dopants and their effect on the graphene electronic properties were purely theoretical.
 Beginning with impurity superlattices, this research preceded the experimental realistation of Nitrogen doped graphene.
By introducing a controlled periodic arrangement of Boron and Nitrogen impurities a band gap was seen to open \cite{casolo_band_2011},
and although such superlattice structures were not feasible on large-scale it was further shown that a random distribution on one sublattice
can also open a gap \cite{botellomendez_modeling_2013}.
Further DFT studies showed that the band gap will increase with dopant concentration \cite{rani_designing_2012,lherbier_electronic_2013} and that
a dopant level of over 8\%, where impurities are all on the same sublattice, will produce a band gap of around 550meV far surpassing the minimum required for a CMOS \cite{lherbier_electronic_2013,kim_role_2011}
and finding that the band gap scales with concentration to the power $3/4$, as shown in Fig \ref{fig:bandgaps}.

Even with a 4:1 doping ratio between sublattices, the band gap, although smaller, was shown to still exist. 
This is promising for scaleability where perfect asymmetric doping may not always be realizable.
Beyond the realization of quasi-ballistic electron transport, sublattice segregated systems can also be used to induce magnetism \cite{nair2012spin, palacios2008vacancy, singh2009magnetism, santos2010magnetism} and produce spin-polarized current \cite{park_spin-polarized_2013, rakyta_effect_2011}.

\begin{figure}
  \centering
  \includegraphics[scale=0.4]{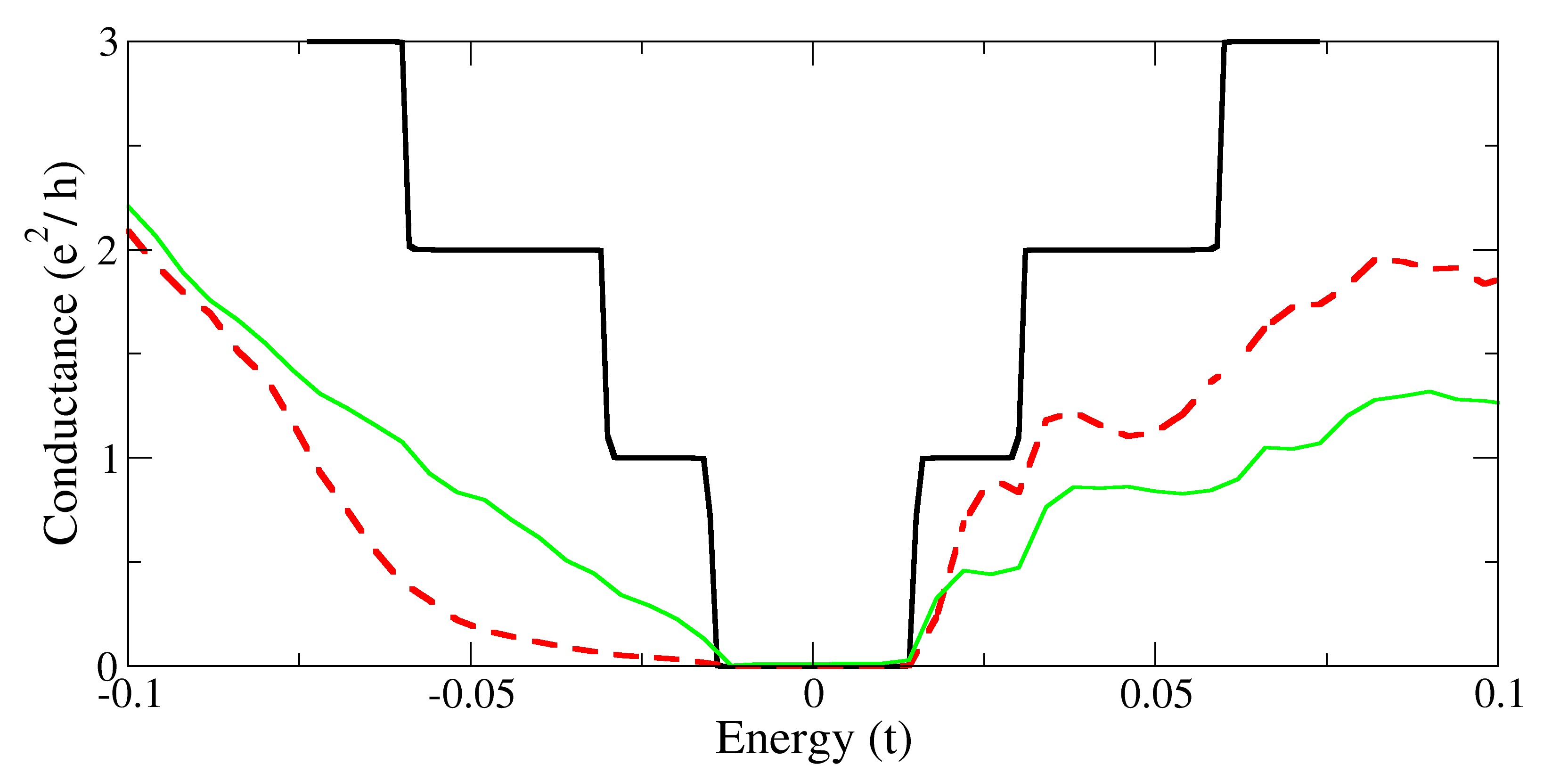}
  \caption{ Quantum conductance through a 15nm wide graphene nanoribbon with a 7.5nm long scattering region containing a dispersion of substitutional nitrogen impurities,
  in a similar vein to the method of Botello-Mendez et al. \cite{botellomendez_modeling_2013},
  calculated using a recursive Green's Function method \cite{economou1984green,sancho1984quick}, the Kubo formula for conductance \cite{kubo1956general} 
  and a configurational average of 50 systems. Energy is in units of the tight binding nearest neighbour hopping energy between carbon atoms, $t = 2.7$eV.
  Shown is the predicted conductance for pristine (black), randomly doped (green solid) and single sublattice doped (red dashed) systems, where a dopant concentration of 1\% nitrogen was used.}
  \label{fig:conductance}
 \end{figure}

Much work has been done studying how the placement of dopants affects the properties of nanoribbons. DFT approaches, using a periodic system of dopants \cite{owens_electronic_2013, chen_semiconductor_2014},
 and a more general Kubo-Greenwood approach \cite{botellomendez_modeling_2013} have shown that electron transport is enhanced when dopants are placed on one sublattice, compared to a random distribution,
 and that a band gap does indeed open. 
 The difference in transport qualities between the asymmetrically doped versus completely randomly doped systems is illustrated by a conductance plot in Fig. \ref{fig:conductance}, 
 where it is evident that the electrons in the former kind of device will be subjected to less scattering leading to an increase in the quantum conductance, closer to that of the pristine system
 when subjected to a positive bias.
 
It should also be noted that that symmetry breaking in nanoribbons occurs via edge effects and also results in a favourable sublattice \cite{power_model_2009}, however the mechanism is distinctly different
from that in graphene. 


\subsection{Outlook}

At the time of writing, the synthesis of sublattice asymmetric graphene is only just now becoming practical. 
There are still many unanswered questions and the most important ones we outline in this section.

The natural question to ask is if this asymmetry is particular to Nitrogen only, or if it can be found with other dopants. 
Theoretical findings in the past have shown that such an effect can be expected with a dilute concentration of certain adsorbates \cite{cheianov_sublattice_2010},
and indeed more recent studies suggest that this could also be possible with other impurities \cite{lawlor_sublattice_2014}. 
 Whilst it is known experimentally that Molybdenum impurities exhibit same sublattice configurations in bilayer epitaxial graphene, 
the mechanism behind this is not currently understood \cite{wan2013incorporating}.
 Boron has been studied in a similar CVD growing regime to Nitrogen where it has been proposed that the role of the substrate can play a critical role in the manifestation of segregation.
 In this case, it is thought that the strong interactions between the Cu(111) crystal and the Boron dopants destroy any asymmetry effects \cite{zhao_local_2013}.
 It is then logical to ask whether suitable substrates can be identified to produce segregated Boron doped graphene sheets, which would have the effect of p-doping the system, and
 then to ask whether this can be extended to other species of dopants.
 We note that it is known that there are weak interactions between graphene and Al, Ag, Au, Pt(111) substrates, all of which leave the electronic structure intact \cite{khomyakov_first-principles_2009},
 whilst substrates such as Ni have considerably stronger interactions \cite{xu_interface_2010}. 
  Another way to investigate the presence of sublattice asymmetry with dopants other than nitrogen is via ion bombardement \cite{wang2011doping}, 
 which could be combined with the high temperature annealing process discussed previously.
 
 The experimental realisation of spin-polarized transport should also be pursued, and could spark additional interest in this research area beyond quasi-ballistic transport.
Investigation of the effects of strain on such a system is difficult due to the technicalities of the CVD method, but could be explored using adsorbates instead
 of substitutions or by incorporation of the strain in to the nitrogen ion bombardement and annealing procedure discussed earlier.
 Another open question is if a critical concentration of dopants exists, as mentioned in the theory section, 
 which would limit the available bandgap below the threshold needed for future use in GFET devices. 
 Currently all experimental reports on sublattice asymmetry have very low concentrations of dopants around 0.3\%, so 
 it should be feasible to test the existence of a critical concentration of using available methods \cite{wang_review_2012}.
 Consequentially this would shed more light on the mechanism responsible for the segregation,
 and whether in fact the inter-impurity interactions and edge growth effects are complementary effects.

\section{Conclusion}

This paper reviews the current state of experimental and theoretical research in sublattice asymmetric Nitrogen doped graphene.
Not only are such systems now able to be synthesised in the lab, we have seen that this area shows promise attaining graphene-based FET devices due to the opening of a bandgap 
whilst maintaining the excellent transport properties of graphene, something which is not realisable in graphene with no sublattice imbalance in dopant distribution.
While there are still many open questions in the field these should be answerable within the scope of current techniques.

\section{Acknowledgements}
The authors acknowledge financial support from the Programme for Research in Third
Level Institutions (PRTLI), Science Foundation Ireland (Grant No. SFI 11/ RFP.1/MTR/3083).

\bibliographystyle{plain}
\bibliography{zotero_bib}

\end{document}